\documentclass[12pt]{article}
\def\cA{{\cal A}}\def\cH{{\cal H}}
\def\bfD{{\bf D}}\def\bfJ{{\bf J}}\def\cT{{\cal T}}
\def\eps{\epsilon}\def\bfI{{\bf I}}\def\been{{\bf 1}}
\begin{document}
\title{On the finite spectral triple of an almost-commutative geometry\footnote{To appear in vol 42 of the Brazilian Journal of Physics.}}
\author
{F.J.Vanhecke\footnote{Instituto de F\'{\i}sica, e-mail : vanhecke@if.ufrj.br} , A.R.da Silva\footnote{Instituto de Matem\'atica, e-mail : beto@im.ufrj.br} and C.Sigaud\footnote{Instituto de F\'{\i}sica : e-mail : sigaud@if.ufrj.br}\\
Universidade Federal do Rio de Janeiro,Brasil}          
\date{}          
\maketitle
\abstract{In this short communication, we examine the relevance of the signature of the space-time metric in the 
construction of the product of a pseudo-Riemannian spectral triple with a finite triple describing the internal geometry.
We obtain arguments favouring the appearance of ${\bf SU}(2)$ and ${\bf U}(1)$ as gauge groups in the standard model.} 
\section{Almost-commutative geometry and the Standard Model of Particle Physics}\label{SM}
The concept of a non-commutative geometry {\bf NCG}, as applied to particle physics, was first proposed by A.Connes in \cite{Alain}. 
A readable account is found in \cite{Joseph}. The main point is that the {\bf NCG} describes the standard model {\bf SM} of elementary particle physics, at least at the Lagrangian level, as the tensor product of two real spectral triples : \\
{\bf 1)} The commutative spectral triple associated to the classical geometry of a Riemannian spin mannifold "Euclidean space-time" with its commutative algebra of functions $\cA_1$, its Clifford structure with self-adjoint Dirac operator $\bfD_1$, hermitian chirality $\Omega_1$ with ${\Omega_1}^2={\bf Id}_1$ and anti-unitary charge conjugation or real structure $\bfJ_1$, all represented on the bona-fide Hilbert space $\cH_1$ of square-integrable spinors.\\
{\bf 2)} A finite spectral triple describing the non-commutative algebra corresponding to the internal quantum numbers. This algebra $\cA_2$ is a direct sum of matrix algebras over the real associative division algebras. It acts on a finite dimensional module $\cH_2$, with a scalar product, an hermitian Dirac operator $\bfD_2$ and chirality $\Omega_2$. It is also endowed with a real structure $\bfJ_2$, corresponding to charge conjugation.\\
This product of a commutative spectral triple with a non-commutative one is termed an 
"almost-commutative geometry" {\bf ACG}. It was mainly plagued by two problems :\\
{\bf I)} : There is a fermion quadruple overcounting \cite{Leonardo}, \cite{Pepe}. \\
{\bf II)} : No neutrino mixing and no neutrino masses are allowed.\\
The classification of the real finite spectral triples was studied in \cite{K} and \cite{PS} and, more recently in \cite{Cacique}.
The real structures are classified by the so-called KO-dimension or signature. This is an integer defined modulo eight which corresponds to the eightfold periodicity of the real Clifford algebras as established in \cite{ABSh}. When the space-time geometry is Lorentzian, not all of the axiomatics of Connes' {\bf NCG} are met. The main difference is  that the space of square integrable spinors does not form a Hilbert space, but at most a Kre\u{\i}n space with indefinite scalar product. 
The Clifford algebra ${\cal C}\ell(\eta)$, associated with a quadratic non-degenerate symmetric form $\eta$, is the real algebra generated by $n$ elements 
$\{\Gamma^\alpha\,;\,\alpha=1,2,\cdots,n\}$ and a unit $\b1$, such that the following relations hold :
$\Gamma^\alpha\,\Gamma^\beta+\Gamma^\beta\,\Gamma^\alpha=+
2\,\eta^{\,\alpha\beta}\,\b1$.
When the metric tensor has $p$ positive and $q$ negative eigenvalues, $p+q=n$, the Clifford algebra is denoted by ${\cal C}\ell(p,q)$. This is Bourbaki's convention  differing from that of \cite{ABSh,Connes}. 
The signature is defined as $\sigma=p-q$ modulo eight.
Besides this signature dimension, the {\bf NCG} has also a metric dimension, which in case of spin manifolds,  coincide with the usual geometric dimension $n=p+q$. In the case of a finite spectral triple with metric dimension $0$, the signature is not necessary zero. Using this fact, A.Connes \cite{Connes} and independently J.Barrett \cite{Barrett} could remedy these shortcomings {\bf I)} and {\bf II)}, allowing the signature of the finite spectral triple to be different from zero. 
Having in mind the relevance of this signature concept, we observe that, in general, the real Clifford algebra ${\cal C}\ell(p,q)$ is different from ${\cal C}\ell(q,p)$ which has the same dimension $n=p+q$.  
Although, for even $n$, both have a unique, up to equivalence, representation in a complex $2^n$-dimensional spinor space ${\cal S}$, they are different real sub-algebras of the complex matrix algebra. 
The associated real structures {\bf J}, generalizing the charge conjugation operator, are also characteristic of each $(p,q)$ choice. The even subalgebras are isomorphic 
${\cal C}\ell^+(p,q)\cong{\cal C}\ell^+(q,p)$ and so are the {\bf SPIN}-groups, ${\bf SPIN}(p,q)\cong{\bf SPIN}(q,p)$, 
covering the special orthogonal group ${\bf SO}(p,q)\cong{\bf SO}(q,p)$.
However the {\bf PIN}-groups are not. Each of ${\bf PIN}(p,q)$ and ${\bf PIN}(q,p)$ are distinct coverings of the isomorphic orthogonal groups ${\bf O}(p,q)\cong{\bf O}(q,p)$. 
This was already observed by Yang and Tiomno \cite{Yang}, with relevant comments in \cite{JR}. It implies a distinct (s)pinor field for a different signature $\sigma$ which changes sign together with the metric. More recently DeWitt-Morette et al.\cite{Berg,Car,Cecile} addressed these issues. 
In \cite{Berg} they examined the double beta decay without neutrinos and conjectured the neutrino to be a Majorana $(3,1)$ fermion. 
In \cite{Car} it was shown, in superstring theory, that the contribution to a Polyakov path integral over Riemannian surfaces is different for a positive metric from that of a negative metric. Finally, in \cite{Cecile} it appeared that for a non trivial topology of configuration space, the fermionic current is different for opposite metrics. \\
The aim of our study is to investigate if the difference between ${\cal C}\ell(p,q)$  and ${\cal C}\ell(q,p)$ is, or is not, relevant in the almost-commutative model of the theory of fundamental interactions.\\
If the KO-dimension is physically significant, so will be the sign of the metric. The physically relevant case is the Minkowski space-time with two possibilities:
${\cal C}\ell(1,3)\cong{\bf M}_2({\bf H})$ with $\sigma=-2=+6\,{\bf mod\,8}$ and 
${\cal C}\ell(3,1)\cong{\bf M}_4({\bf R})$ with $\sigma=+2$.
Their even subalgebras are isomorphic ${\cal C}\ell^+(1,3)\cong{\cal C}\ell^+(3,1)\cong{\bf M}_2({\bf C})$. 
As for an Euclidean space-time, there is only one ${\cal C}\ell(4,0)\cong{\cal C}\ell(0,4)\cong{\bf M}_2({\bf H})$
with $\sigma=+4=-4\,{\bf mod\,8}$ and the even subalgebra ${\cal C}\ell^+(4,0)\cong{\cal C}\ell^+(0,4)\cong{\bf H}\oplus{\bf H}$.\\
The signature of a real, even, spectral triple is determined by the three $\pm1$-valued 
numbers $\{\epsilon,\,\epsilon^\prime,\,\epsilon^{\prime\prime}\}$, given in table {\bf\ref{Tafel}.} and defined by :
\[\bfJ^2=\epsilon\, {\bf Id}\;,\;\bfJ\,\bfD=\epsilon^\prime\,\bfD\,\bfJ\;,\;\bfJ\,\Omega=\epsilon^{\prime\prime}\,\Omega\,\bfJ\]
For odd dimensions, there is no chirality available to define the $\epsilon^{\prime\prime}$ sign. Consequently the corresponding entry is left blank. As for even dimensional triples, the {\it volume element} $\Theta\doteq\Gamma^1\,\Gamma^2\cdots\Gamma^n$ is unitary and has square $\Theta^2=(-1)^{\sigma/2}$. For $\sigma=0,4$ we define $\Omega=\Theta$, which is hermitian and belongs to the real Clifford algebra. When $\sigma=2,6$, the volume form squares to $-\,\been$ and, in order to have states of definite chirality we must complexify the representation space and 
$\Omega\doteq\imath\,\Theta$ is an hermitian operator in representation space but, obviously, does not belong to the real Clifford algebra. 
In earlier work \cite{Chicus,Goyi} we pointed out some difficulties in the very definition of the tensor product of real spectral triples. The calculations in this work are used to solve the following problem : 
{\it Given an even spectral triple ${\cal T}_1$, what are the possible spectral triples ${\cal T}_2$ such that their product, $\cT=\cT_1\otimes\cT_2$, is an even spectral triple with certain required properties.}
\begin{table}\caption{The epsilon Table}\label{Tafel}
\begin{center}
\begin{tabular}{|c||c|c|c|c|c|c|c|c|} \hline
$\sigma=$                & 0  & 1  & 2 & 3  & 4  & 5  & 6  &  7 \\ \hline\hline
$\epsilon$               & +1 & +1 & +1& -1 & -1 & -1 & -1 & +1 \\ \hline
$\epsilon^\prime$        & +1 & +1 & +1& -1 & +1 & +1 & +1 & -1 \\ \hline
$\epsilon^{\prime\prime}$& +1 &    & -1&    & +1 &    & -1 &    \\ \hline
\end{tabular}\end{center}
\end{table}
\section{Defining the Product $\cT_1\otimes\cT_2$}\label{Het Produkt}
In this section we strive to define the tensor product of two even spectral triples $\cT_1=\{\cA_1,\cH_1,\bfD_1,\Omega_1,\bfJ_1\}$ and $\cT_2=\{\cA_2,\cH_2,\bfD_2,\Omega_2,\bfJ_2\}$, given by  
$\cT=\cT_1\otimes\cT_2=\{\cA,\cH,\bf D,\Omega,\bfJ\}$ with :
\begin{eqnarray*}
&&
\cA=\cA_1\otimes\cA_2\;,\;\cH=\cH_1\otimes\cH_2\;,\\
&&
\bfD=\bfD_1\otimes\bfI_2+\Omega_1\otimes\bfD_2\;,\;
\Omega=\Omega_1\otimes\Omega_2\end{eqnarray*}
The Dirac operator $\bfD$ and the chirality $\Omega$ are constructed such that the resulting Dirac operator is odd : $\bfD\,\Omega+\Omega\,\bfD=0$, and such that the metric dimensions add : $n=n_1+n_2$ since $\bfD^2={\bfD_1}^2+{\bfD_2}^2$. In order to define the real structure of the product, we attempt a first {\it"natural choice} :
\begin{equation}\label{Nat1}\bfJ=\bfJ_1\otimes\bfJ_2\end{equation}
\subsection{The natural real structure $\bfJ=\bfJ_1\otimes\bfJ_2$}\label{Verwacht}
We compute 
\begin{eqnarray*}
\bfJ^2&=&
\eps_1\;\eps_2\,\bfI=\eps\,\bfI\\
\bfJ\,\bfD&=&
\left(\eps^\prime_1\,\bfD_1\otimes\bfI_2+\eps^{\prime\prime}_1\,\eps^\prime_2\,\Omega_1\otimes\bfD_2\right)\,(\bfJ_1\otimes\bfJ_2)=\eps^\prime\,\bfD\,\bfJ\\
\bfJ\,\Omega&=&
\eps^{\prime\prime}_1\,\eps^{\prime\prime}_2\,\Omega\,\bfJ=
\eps^{\prime\prime}\,\Omega\,\bfJ
\end{eqnarray*}
which are consistent if we require that 
\begin{equation}\label{Nat2}
\eps=\eps_1\,\eps_2\;,\;
\eps^\prime=\eps^\prime_1=\eps^{\prime\prime}_1\,\eps^\prime_2\;,\;
\eps^{\prime\prime}=\eps^{\prime\prime}_1\,\eps^{\prime\prime}_2\end{equation}
The signatures of the spectral triples will be denoted respectively by $\sigma_1\,,\,\sigma_2$ and $\sigma$. Consider the different possibilities for $\sigma_1$ :
\[{\bf 1)}\quad\sigma_1=0\;;\;
\eps=+\,\eps_2\;,\;
\eps^\prime=+1=+\,\eps^\prime_2\;,\;
\eps^{\prime\prime}=+\,\eps^{\prime\prime}_2\] 
It is seen that any even value of the spectral dimension of the second factor is possible.
If chirality of the second factor is irrelevant, the values $\sigma_2\in\{1,5\}$ are also allowed.
\[{\bf 2)}\quad\sigma_1=2\;;\;
\eps=+\,\eps_2\;,\;\eps^\prime=+1=-\,\eps^\prime_2\;,\;
\eps^{\prime\prime}=-\,\eps^{\prime\prime}_2\]
No even value of $\sigma_2$ is possible, but $\sigma_2\in\{3,7\}$ is allowed.
\[{\bf 3)}\quad\sigma_1=4\;;\;
\eps=-\,\eps_2\;,\;
\eps^\prime=+1=+\,\eps^\prime_2\;,\;
\eps^{\prime\prime}=+\,\eps^{\prime\prime}_2\]
All even values of $\sigma_2$ are allowed and also, without chirality, the odd values $\sigma_2\in\{1,5\}$.
\[{\bf 4)}\quad\sigma_1=6\;;\;\eps=+\,\eps_2\;,\;
\eps^\prime=+1=-\,\eps^\prime_2\;,\;
\eps^{\prime\prime}=-\,\eps^{\prime\prime}_2\] 
Again no even $\sigma_2$ values but only $\sigma_2\in\{1,5\}$ are allowed. \\
Observe that, in each case, the signature is additive modulo eight :
\begin{equation}\label{Nat3}\sigma=\sigma_1+\sigma_2\end{equation}
Connes \cite{Connes} requires the KO-dimension of the product to be $\sigma=6$ with commutation relations $\bfJ^2=-\bfI$, $\bfJ\,\bfD=\bfD\,\bfJ$ and $\bfJ\,\Omega=-\,\Omega\,\bfJ$. Since its first factor is Euclidean, $\sigma_1=4$, according to \ref{Nat3}, the second factor must be a $\sigma_2=2$.
\subsection{The modified real structure $\bfJ=\bfJ_1\otimes\bfJ_2\,\Omega_2$ }\label{Oplossing}
From {\bf \ref{Verwacht}} it is seen that for a Minkowski signature $\sigma_1\in\{2,6\}$, the product can only be defined with a second odd factor and the chirality paradigm is lost.
This problem was already noticed in \cite{Chicus}. If we have two even factors, another choice for $\bfJ$ is provided by a modified tensor product of the real structures :
\begin{equation}\label{Nat4}\bfJ=\bfJ_1\otimes\bfJ_2\,\Omega_2\end{equation}
Again, we compute 
\begin{eqnarray*}
\bfJ^2&=&
\eps_1\;\eps_2\,\eps^{\prime\prime}_2\,\bfI=\eps\,\bfI\\
\bfJ\,\bfD&=&
\left(\eps^\prime_1\,\bfD_1\otimes\bfI_2-\eps^{\prime\prime}_1\,\eps^\prime_2\,\Omega_1\otimes\bfD_2\right)\,(\bfJ_1\otimes\bfJ_2\,\Omega_2)=\eps^\prime\,\bfD\,\bfJ\\
\bfJ\,\Omega&=&
\eps^{\prime\prime}_1\,\eps^{\prime\prime}_2\,\Omega\,\bfJ=
\eps^{\prime\prime}\,\Omega\,\bfJ
\end{eqnarray*}
and, for consistency, we require that 
\begin{equation}\label{Nat5}
\eps=\eps_1\,\eps_2\,\eps^{\prime\prime}_2\;,\;
\eps^\prime=\eps^\prime_1=-\,\eps^{\prime\prime}_1\,\eps^\prime_2\;,\;
\eps^{\prime\prime}=\eps^{\prime\prime}_1\,\eps^{\prime\prime}_2\end{equation}
Again we consider the different possibilities $\sigma_1$ :
\[{\bf 1)}\quad\sigma_1=0\;;\;
\eps=+\eps_2\,\eps^{\prime\prime}_2\;,\;
\eps^\prime=+1=-\,\eps^\prime_2\;,\;
\eps^{\prime\prime}=+\eps^{\prime\prime}_2\quad\]
Since $\eps^\prime_2=-1$, only odd values of $\sigma_2$ are allowed. Furthermore, no chirality $\Omega_2$ is available to define $\bfJ$. 
 \[{\bf 2)}\quad\sigma_1=2\;;\;
\eps=-\,\eps_2\,\eps^{\prime\prime}_2\;,\;
\eps^\prime=+1=+\,\eps^\prime_2\;,\;
\eps^{\prime\prime}=-\,\eps^{\prime\prime}_2\]
Now all even values of $\sigma_2$ are allowed and again $\sigma=\sigma_1+\sigma_2$.
In particular, if $\sigma_2=6$ we obtain $\sigma=0$.
\[{\bf 3)}\quad\sigma_1=4\;;\;
\eps=-\,\eps_2\,\eps^{\prime\prime}_2\;,\;
\eps^\prime=+1=-\,\eps^\prime_2\;,\;
\eps^{\prime\prime}=+\eps^{\prime\prime}_2\]
Again the second triple must be odd and $\bfJ$ is not defined.
\[{\it 4)}\quad\sigma_1=6\;;\;
\eps=+\eps_2\,\eps^{\prime\prime}_2\;,\;
\eps^\prime=+1=+\,\eps^\prime_2\;,\;
\eps^{\prime\prime}=-\,\eps^{\prime\prime}_2\]
All even values are allowed. The additivity $\sigma=\sigma_1+\sigma_2$ still holds.\\
Barrett \cite{Barrett} requires a product with $\sigma=0$ such that 
$\bfJ^2=+\bfI$, $\bfJ\,\bfD=\bfD\,\bfJ$ and $\bfJ\,\Omega=+\,\Omega\,\bfJ$. 
If we have a Minkowski space-time respectively of signature (modulo 8) $\sigma_1=+2$ or $\sigma_1=6$, the finite triple should have signature $\sigma_2=6$ or $\sigma_2=+2$. 
We may then restrict the representation space to the common eigenstates of $\bfJ$ and $\Omega$ 
with eigenvalues $\pm 1$. This would cure the problem {\bf I)} of overcounting. Since there are two different kinds of spinor fields, each with a different internal symmetry, the space of states is increased. 
\noindent
\section{Summary and perspectives}\label{Besluit}
We examined possible consistent products of a (pseudo)Riemannian spectral triple with a finite spectral triple. In the Minkowski case the signature may be $\sigma_1\in\{+2,-2\}$ and it was seen that the possible signatures of the finite triple are given, respectively, as $\sigma_2\in\{-2,+2\}$. This implies the possibility of having two kind of spinors each with a different internal algebra. 
The ${\cal C}\ell(3,1)$ fermions of given chirality should couple to ${\cal C}\ell(0,2)\cong{\bf H}$ with unitary group ${\bf SU}(2)$. On the other hand, a ${\cal C}\ell(1,3)$ fermion of opposite chirality couples to ${\cal C}\ell(2,0)\cong{\bf M}_2({\bf R})$ with unitary group ${\bf SO}(2)\cong{\bf U}(1)$. \\
In order to construct the required finite triple we could take a Riemannian two-dimensional triple of the required signature and, inspired by Kaluza-Klein theory, collapse its base manifold to a point.  This can be implemented through the substitution of the partial derivatives $\partial/\partial x^\mu$ by a constant vector. However, such a procedure shows to be incompatible with some of the required properties of a real spectral triple. In particular the self-adjointness of the Dirac operator conflicts with its behaviour under the anti-unitary charge conjugation $\bfJ$.     Fortunately a solution is provided if one realizes that, in the massless case, two possibilities are open with the charge conjugation operator, namely $\bfJ_\pm$ such that $\bfJ_\pm\,\bfD=\pm\,\bfD\,\bfJ_\pm$. 
They are related by $\bfJ_-=\bfJ_+\,\Omega$, a modified real structure that appeared also in \cite{Chicus}. Such generalization of charge conjugation has also been considered in the framework of supergravity \cite{Casanova}. Also it was recently examined in \cite{DabeDoss} and provides a generalization of the tensor product construction of \cite{Chicus}. In forthcoming work, we attempt to construct explicitly finite real spectral triples of this kind. The Dirac operator must also be generalized with a gauge field what, in Connes' terminology, provides a fluctuation of the metric. 
Possible Higgs mechanism generating mass, should implement a toy model of electroweak interactions.\\
 Finally, we would like to mention the work, started with \cite{Lou} and elaborated on in \cite{Mir}, to relate ${\cal C}\ell(1,3)$ with ${\cal C}\ell(3,1)$, showing the relevance of the challenge : real versus complex Clifford algebras.

\end{document}